\documentclass[prb,aps,twocolumn,showpacs]{revtex4-2} 
\usepackage{amsmath,amssymb,amsthm}
\usepackage{graphicx}
\usepackage{subfigure}
\usepackage{amsmath}
\usepackage{amsfonts,amssymb}
\usepackage{bm}
\usepackage{float}
\usepackage{color}
\usepackage{sidecap}
\allowdisplaybreaks
\usepackage{soul}
\setstcolor{red}
\usepackage[normalem]{ulem}
\usepackage{hyperref}
\hypersetup{colorlinks=true,breaklinks,urlcolor=blue,linkcolor=blue,citecolor=blue}

\begin{document}

\title{Topological superconducting phase in a non-Hermitian Kitaev chain with staggered pairing imbalance}

\author{Xiao-Jue Zhang$^{1}$}
\author{Rong L\"u$^{2,3}$}
\author{Qi-Bo Zeng$^{1}$}
\email{zengqibo@cnu.edu.cn}
\affiliation{$^{1}$Department of Physics, Capital Normal University, Beijing 100048, China}
\affiliation{$^{2}$State Key Laboratory of Low-Dimensional Quantum Physics, Department of Physics, Tsinghua University, Beijing 100084, China}
\affiliation{$^{3}$Collaborative Innovation Center of Quantum Matter, Beijing 100084, China}

\begin{abstract}
We introduce a one-dimensional non-Hermitian Kitaev chain with staggered imbalance in the $p$-wave superconducting pairing. By tuning the chemical potential and the pairing imbalance, we find that the eigenenergy spectrum undergoes real-to-complex transitions, and the spectral gap can change from a real line gap to an imaginary one. The pairing imbalance significantly enlarges the parameter region supporting a topological superconducting phase. Remarkably, we show that a topologically nontrivial phase hosting Majorana zero modes can be induced by varying the pairing imbalance, even in the regime of strong chemical potential. The gap-closing points and phase boundaries are determined analytically, and the resulting phase diagrams are presented with the nontrivial phase characterized by a nonzero topological invariant. Furthermore, we identify the coexistence of Majorana zero modes and finite-energy Majorana edge modes in this system. Our results reveal exotic phenomena arising from imbalanced pairing and establish a platform for exploring topological superconductivity in non-Hermitian systems.
\end{abstract}
\maketitle
\date{today}

\section{Introduction}
Topological superconductors (TSCs) have been extensively investigated over the past two decades as promising platforms for realizing Majorana fermions \cite{Alicea2012PRP,Beenakker2013ARCMP,Elliott2015RMP,Ando2015ARCMP,Sato2017RPP,Yazdani2023Science}. In their topologically nontrivial phases, TSCs host Majorana zero modes (MZMs) localized at system boundaries, which obey non-Abelian statistics and are therefore of central interest for fault-tolerant quantum computation based on braiding operations \cite{Alicea2011NatPhys,Sarma2015npj,Obrien2018PRL,Lian2018PNAS,Litinski2018PRB}. A wide variety of theoretical proposals have been put forward to realize MZMs, including two-dimensional p-wave superconductors~\cite{Read2000PRB,Stone2004PRB,Fendley2007PRB}, superconductor heterostructures coupled to topological insulators \cite{Fu2008PRL} or semiconductors \cite{Sau2010PRL,Sau2010PRB,Lutchyn2010PRL,Oreg2010PRL}, as well as other solid-state platforms \cite{Chung2011PRB,Raghu2010PRL}. In addition, TSC phases have been explored in superfluid 
He-3 \cite{Kopnin1991PRB,Qi2009PRL,Chung2009PRL}, ultracold atomic gases \cite{Zhang2008PRL,Sato2009PRL,Liu2012PRA,Qu2013NatCom,Chen2013PRL,Qu2015PRA,Ruhman2015PRL}, and chains of magnetic atoms on superconducting substrates \cite{Perge2013PRB,Hui2015SR,Dumitrescu2015PRB}. Among these platforms, one-dimensional semiconductor–superconductor heterostructures have attracted particular attention \cite{Sau2010PRB,Lutchyn2010PRL,Oreg2010PRL}, and signatures of MZMs in such systems have been actively investigated both theoretically \cite{Ioselevich2011PRL,Zazunov2011PRB,Wu2012PRB,Jose2012PRL,Ueda2014PRB,Zeng2016FP} and experimentally \cite{Mourik2012Science,Deng2012NanoLett,Rokhinson2012NatPhys,Das2012NatPhys,Finck2013PRL,Perge2014Science,Cao2023SciChina,Kouwenhoven2025MPLB}.

A paradigmatic lattice model for one-dimensional topological superconductivity is the Kitaev chain proposed in Ref.~\cite{Kitaev2001}. Since then, numerous extensions of the Kitaev model have been explored. These include studies of periodic, quasiperiodic, and disordered potentials and their effects on MZMs \cite{Akhmerov2011PRL,Cai2013PRL,DeGottardi2013PRL,Wang2016PRB,Zeng2016PRB,Zeng2021EPL}, Kitaev chains with long-range hopping and/or pairing \cite{Vodola2014PRL,Antonio2017PRB,Dutta2017PRB,Fraxanet2022PRB,Francica2022PRB,Huang2024PRB}, and dimerized Kitaev chain models~\cite{Wakatsuki20142014PRB,Wang2017PRB,Ezawa2017PRB,Roy2023PRB,Roy2024SciRep}. In addition, generalized Kitaev models with modulated 
$p$-wave pairing and hopping have been studied in both one and two dimensions \cite{Liu2016CPB,Zhou2016CPB,Zhou2017PLA,Lesser2020PRR}. Furthermore, inhomogeneous or periodically varying superconductivity has been shown to substantially modify the topological properties of Kitaev chains \cite{Hoffman2016PRB,Levine2017PRB,Escribano2019PRB,Zhang2025JPCM}. In parallel, non-Hermitian systems have attracted considerable attention as effective descriptions of open quantum systems in recent years \cite{Cao2015RMP,Konotop2016RMP,Ashida2020AP,Bergholtz2021RMP}. The presence of non-Hermiticity can give rise to phenomena that are absent in Hermitian settings, such as exceptional points \cite{Heiss2012JPA} and the non-Hermitian skin effect \cite{Yao2018PRL1,Yao2018PRL2}, which profoundly alters the nature of topological phases, including possible breakdowns of bulk–boundary correspondence~\cite{Kawabata2019PRX}. Motivated by these developments, the studies on TSCs and Kitaev chains have also been extended to non-Hermitian regimes, where the effects of gain and loss on the topological superconductivity and MZMs have been investigated \cite{Wang2015PRA,Zeng2016PRA,Li2018PRB,Shibata2019PRB,Zhao2021PRB,Shi2023PRB1,Shi2023PRB2,Sayyad2023PRR,Ghosh2024SciPost,Cayao2024arxiv,Chang2025arxiv}. In particular, several non-Hermitian Kitaev chain models with imbalanced pair creation and annihilation were shown to exhibit exotic topological properties \cite{Li2018PRB,Zhao2021PRB,Shi2023PRB1,Shi2023PRB2}. It will be interesting to ask if the pairing imbalance becomes spatially modulated, how the topological superconducting phase and MZMs will behave in such non-Hermitian systems.

In this work, we study a one-dimensional non-Hermitian Kitaev chain with staggered imbalance in the superconducting pairing, which describes the pairing imbalances in the odd (even) dimers. By tuning the chemical potential and the pairing imbalance, we find that the eigenenergy spectrum undergoes real-to-complex transitions, and the spectral gap can change from a real to an imaginary line gap. The corresponding transition points and gap-closing conditions are determined analytically from the momentum-space spectrum. We further show that the presence of pairing imbalance substantially enlarges the parameter region supporting a topologically nontrivial superconducting phase compared with the Hermitian limit. Remarkably, due to the interplay between the chemical potential and the staggered pairing imbalance, the topological phase and Majorana zero modes can persist even in the regime of arbitrarily large chemical potential, provided that the imbalances are introduced in an alternating manner. The nontrivial phases are characterized by a nonzero topological invariant, and the resulting phase diagrams are presented. In addition, we identify a regime in which Majorana zero modes coexist with finite-energy Majorana edge modes. Our results demonstrate that non-Hermitian pairing imbalance can play a constructive role in stabilizing topological superconductivity, and also unveil the influences of the real/imaginary line gaps on the topological phases. The model introduced in this work thus provides a versatile platform for exploring topological phases in non-Hermitian superconducting systems.

The remainder of this paper is organized as follows. In Sec.~\ref{Sec2}, we introduce the model Hamiltonian of the one-dimensional Kitaev chain with staggered pairing imbalance. In Sec.~\ref{Sec3}, we analyze the eigenenergy spectrum and the associated line gaps. The topological phases and Majorana edge modes are investigated in Sec.~\ref{Sec4}. Finally, Sec.~\ref{Sec5} summarizes our results.

\section{Model Hamiltonian}\label{Sec2}
\begin{figure}[t]
	\includegraphics[width=3.4in]{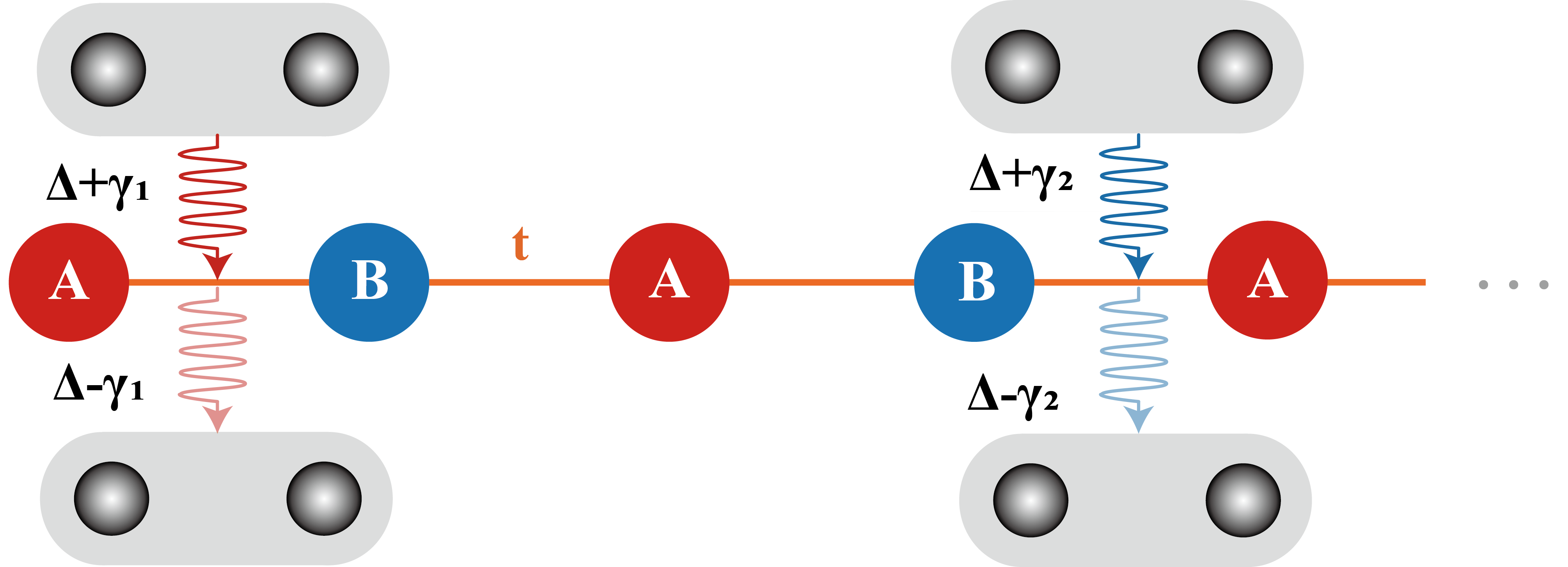}
	\caption{(Color online) Schematic illustration of the one-dimensional Kitaev chain with staggered $p$-wave superconducting pairing imbalance under open boundary conditions. Each unit cell contains two sites, labeled A and B, with a uniform chemical potential $\mu$. The orange lines denote nearest-neighbor hopping with amplitude $t$. The gray shaded regions indicate staggered superconducting pairing terms $\Delta \pm \gamma_{1,2}$ on the intra- and intercell bonds, respectively, which give rise to non-Hermiticity through imbalanced pair creation and annihilation.}
	\label{fig1}
\end{figure}

We consider a one-dimensional lattice model with staggered imbalance in the 
$p$-wave superconducting pairing. A schematic illustration of the lattice under open boundary conditions (OBCs) is shown in Fig.~\ref{fig1}. Owing to the staggered pairing imbalance, each unit cell contains two sublattice sites, labeled A and B. The system is described by the Hamiltonian
\begin{equation}\label{H}
	\begin{aligned}
	H =& -\sum_{j=1}^{N} \mu \left( c_{j,A}^\dagger c_{j,A} + c_{j,B}^\dagger c_{j,B} \right) \\ 
	&-\sum_{j=1}^{N} t \left(  c_{j,B}^\dagger c_{j,A} + h.c. \right) - \sum_{j=1}^{N-1} t \left( c_{j+1,A}^\dagger c_{j,B} + h.c. \right) \\ 
	&+ \sum_{j=1}^{N} \left[ (\Delta +\gamma_1) c_{j,B}^\dagger c_{j,A}^\dagger  + (\Delta -\gamma_1) c_{j,A} c_{j,B} \right] \\
	&+ \sum_{j=1}^{N-1} \left[ (\Delta +\gamma_2) c_{j+1,A}^\dagger c_{j,B}^\dagger + (\Delta -\gamma_2) c_{j,B} c_{j+1,A} \right].
	\end{aligned}
\end{equation}
Here, $c_{j,\alpha}$ ($c_{j,\alpha}^\dagger$) annihilates (creates) a spinless fermion at sublattice site $\alpha=A, B$ in the $j$th unit cell. The total number of unit cells is $N$, corresponding to a chain length $L=2N$. The chemical potential $\mu$ is taken to be uniform throughout the lattice. The parameter $t$ denotes the nearest-neighbor hopping amplitude, and throughout this work we set $t=1$ as the energy unit. The quantity $\Delta$ represents the amplitude of the $p$-wave superconducting pairing, while $\gamma_1$ and $\gamma_2$ are formally introduced to represent the pairing imbalances in the intra- and intercell bonds, respectively. As a result, the Hamiltonian in Eq.~(\ref{H}) is intrinsically non-Hermitian and exhibits properties absent in its Hermitian counterpart.

\subsection{Real-space Bogoliubov–de Gennes formulation}
The eigenenergy spectrum of the Hamiltonian in Eq.~(\ref{H}) under OBC is obtained via exact diagonalization within the Bogoliubov–de Gennes (BdG) formalism. We introduce quasiparticle operators
\begin{equation}
	\eta_{n,\alpha} =  \sum_{j=1}^{N} \left[ u_{n,j \alpha} c_{n,j \alpha}^\dagger + v_{n,j \alpha} c_{n,j \alpha} \right], \alpha \in \left\{ A, B\right\}
\end{equation}
where $n$ labels the eigenstates of the system. The corresponding eigenstates $| \Psi_n \rangle$ satisfy the Schr\"odinger equation $H | \Psi_n \rangle = E_n | \Psi_n \rangle$. Each eigenstate can be represented as a $2L$-component Nambu spinor $| \Psi_n \rangle = \left[u_{n,1A}, u_{n,1B},\cdots,v_{n,jA}, v_{n,jB},\cdots \right]^T$, and the Hamiltonian is cast into a $2L \times 2L$ BdG matrix, whose eigenvalues yield the full spectrum. 

To characterize the localization properties of the eigenstates, we employ the inverse participation ratio (IPR), defined as 
\begin{equation}\label{IPR}
	\text{IPR}_n = \sum_{j,\alpha} \left( |u_{n,j\alpha}|^4 + |v_{n,j\alpha}|^4 \right).
\end{equation}
For extended states, the IPR scales to zero in the thermodynamic limit, whereas it remains finite for localized states.

\subsection{Momentum-space Hamiltonian}
Under periodic boundary conditions (PBCs), the Hamiltonian can be Fourier transformed into momentum space. Introducing the Nambu spinor $C_k^\dagger = \begin{bmatrix} c_{k,A}^\dagger & c_{k,B}^\dagger & c_{-k,A} & c_{-k,B} \end{bmatrix}$, the Hamiltonian takes the form
\begin{equation}
	H = \frac{1}{2} \sum_k C_k^\dagger h(k) C_k ,
\end{equation}
where the BdG Hamiltonian $h(k)$ is given by
\begin{widetext}
\begin{equation}\label{hk}
	h(k) = \begin{bmatrix} 
		-\mu & -t(1+e^{-ik}) & 0 & - (\Delta +\gamma_1)+(\Delta +\gamma_2) e^{-ik} \\
		-t(1+e^{ik}) & -\mu & (\Delta +\gamma_1) - (\Delta +\gamma_2) e^{ik} & 0 \\
		0 & (\Delta -\gamma_1) - (\Delta -\gamma_2) e^{-ik} & \mu & t(1+e^{-ik}) \\
		- (\Delta -\gamma_1)+(\Delta -\gamma_2) e^{ik} & 0 & t(1+e^{ik}) & \mu
	\end{bmatrix}.
\end{equation}
By setting $\gamma_1=\gamma_2=\gamma$, we can recover the Hamiltonian in Ref.~\cite{Li2018PRB}, which studies the non-Hermitian Kitaev chain with uniform pairing imbalance in the system.

The Hamiltonian $h(k)$ exhibits a well-defined symmetry structure characteristic of non-Hermitian BdG systems. Owing to the intrinsic particle–hole redundancy of the BdG representation, it satisfies particle–hole symmetry (PHS), implemented by $\mathcal{C}=\tau_x \mathcal{K}$, such that $\mathcal{C}h(k) \mathcal{C}^{-1}=-h(-k)$, regardless of the non-Hermiticity introduced by the imbalanced pairing terms. Here $\tau_x$ is the Pauli matrix and $\mathcal{K}$ is the conjugate operator and we have $\mathcal{C}^2=+1$. In contrast, time-reversal symmetry (TRS), defined by $\mathcal{T}=\mathcal{K}$ for this spinless system, requires $h^{*}(k)=h(-k)$, which is generally violated when $\gamma_1 \neq \gamma_2$ due to the staggered pairing structure. However, when $\gamma_1 = \gamma_2$, the TRS restores. Therefore, for the general cases with $\gamma_1 \neq \gamma_2$ we study in this work, the system belongs to the $D^\dagger$ class, and can be characterized by a $\mathbb{Z}$ or $\mathbb{Z}_2$ topological invariant depending on whether the energy gap is a point gap or a real line gap, respectively. However, if the gap is an imaginary line gap, the system is trivial~\cite{Kawabata2019PRX}.

Diagonalizing $h(k)$ yields the energy dispersion relation (omitting the overall factor $1/2$),
\begin{equation}\label{Ek}
	\begin{aligned}
	E^2(k) &= \mu^2 - \gamma_1^2 - \gamma_2^2 + 2 \gamma_1 \gamma_2 \cos k + 2 t^2(1+\cos k) + 2 \Delta^2 (1-\cos k) \\ 
	&\pm 2 \sqrt{2t^2 \mu^2 (1+\cos k)-\left[ t^2(1+\cos k)^2 + \Delta^2 (1-\cos^2 k ) \right] (\gamma_1 - \gamma_2)^2}.
	\end{aligned}
\end{equation}
\end{widetext}
From Eq.~(\ref{Ek}), the gap-closing conditions and the phase boundaries between topologically trivial and nontrivial phases can be determined analytically. In the following sections, we analyze the spectral properties of the model and investigate the resulting non-Hermitian topological superconducting phases and Majorana edge modes.

\begin{figure}[t]
	\includegraphics[width=3.4in]{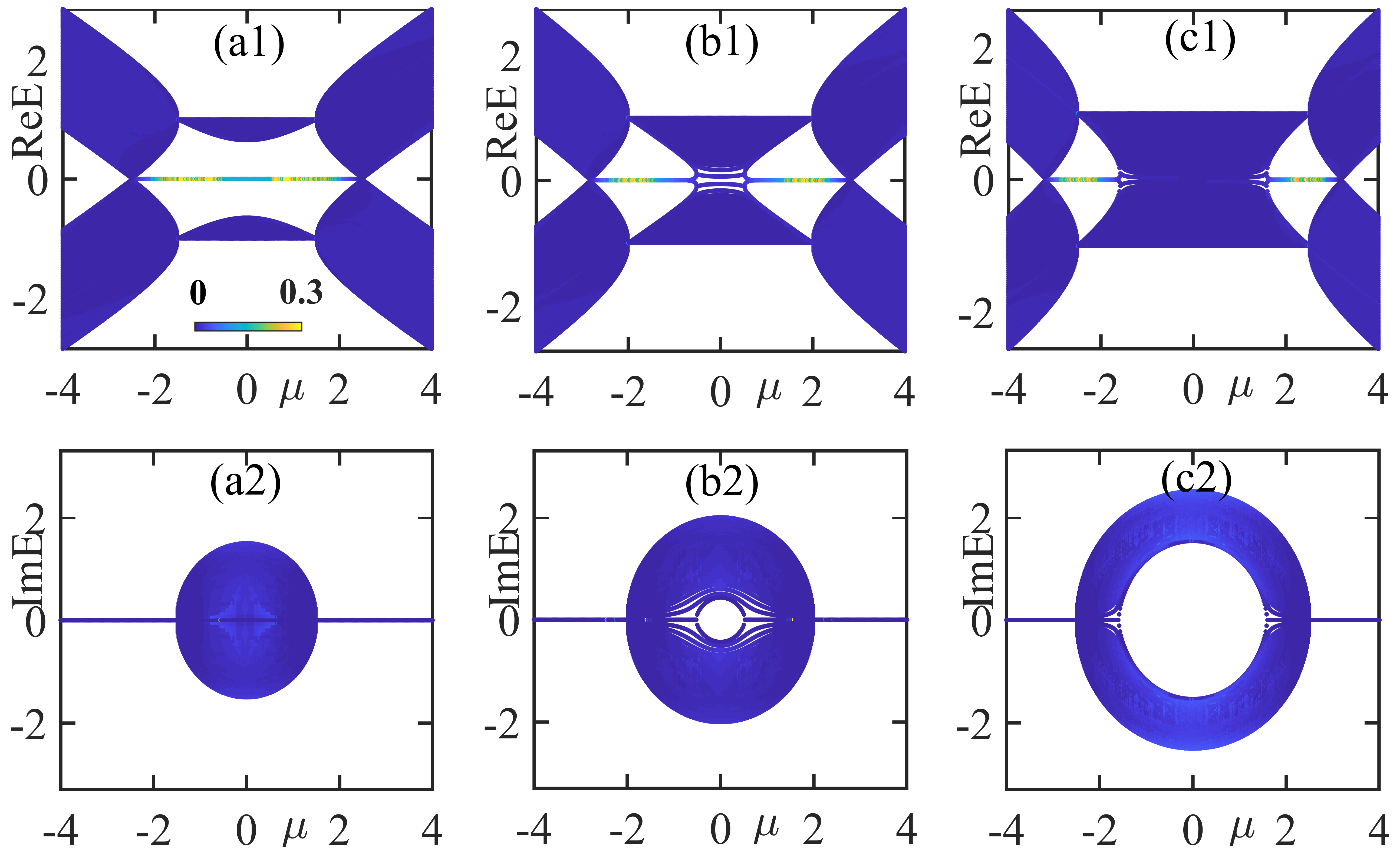}
	\caption{(Color online) Real and imaginary parts of the eigenenergy spectrum as a function of chemical potential $\mu$ for different values of $\gamma_1$ with $\gamma_2=0$. Panels (a)–(c) correspond to (a) $\gamma_1=1.5$, (b) $\gamma_1=2.0$, (c) $\gamma_1=2.5$. The color scale denotes the inverse participation ratio (IPR) of the eigenstates. Other parameters are $t=1$, $\Delta=1$, and system size $L=2N=200$.}
	\label{fig2}
\end{figure}

\section{Eigenenergy spectrum and energy gaps}\label{Sec3}
Non-Hermitian systems normally exhibit various spectral structures that differ significantly from their Hermitian counterparts. The spectrum of such systems often consists of complex eigenvalues, which can be understood through the presence of point gaps, real line gaps, and imaginary line gaps. A point gap exists if the complex-energy bands do not cross a reference point $E=E_p$ in the complex-energy plane. While real (imaginary) line gaps are associated with the separation of real (imaginary) parts of eigenvalues, where the complex-energy bands do not cross a real (imaginary) reference line in the complex-energy plane~\cite{Kawabata2019PRX}. Together, these gaps provide deep insights into the unique behavior and dynamics of non-Hermitian systems. In the following, we will check the spectral structures of the model introduced in this work.

We first analyze the eigenenergy spectrum for the case $\gamma_2=0$, corresponding to pairing imbalance introduced on alternating bonds. Figures~\ref{fig2}(a1) and \ref{fig2}(a2) display the real and imaginary parts of the spectrum as functions of the chemical potential $\mu$, where the color scale denotes the IPR of the eigenstates as defined in Eq.~(\ref{IPR}). For $|\gamma_1|<2t$, the real part of the spectrum exhibits a finite gap, while the imaginary part remains gapless, indicating the presence of a real line gap. The real gap closes at
\begin{equation}
	\mu = \pm \sqrt{4t^2 + (\gamma_1-\gamma_2)^2},
\end{equation}
which follows from Eq.~(\ref{Ek}) by setting $k=0$.

\begin{figure}[t]
	\includegraphics[width=3.4in]{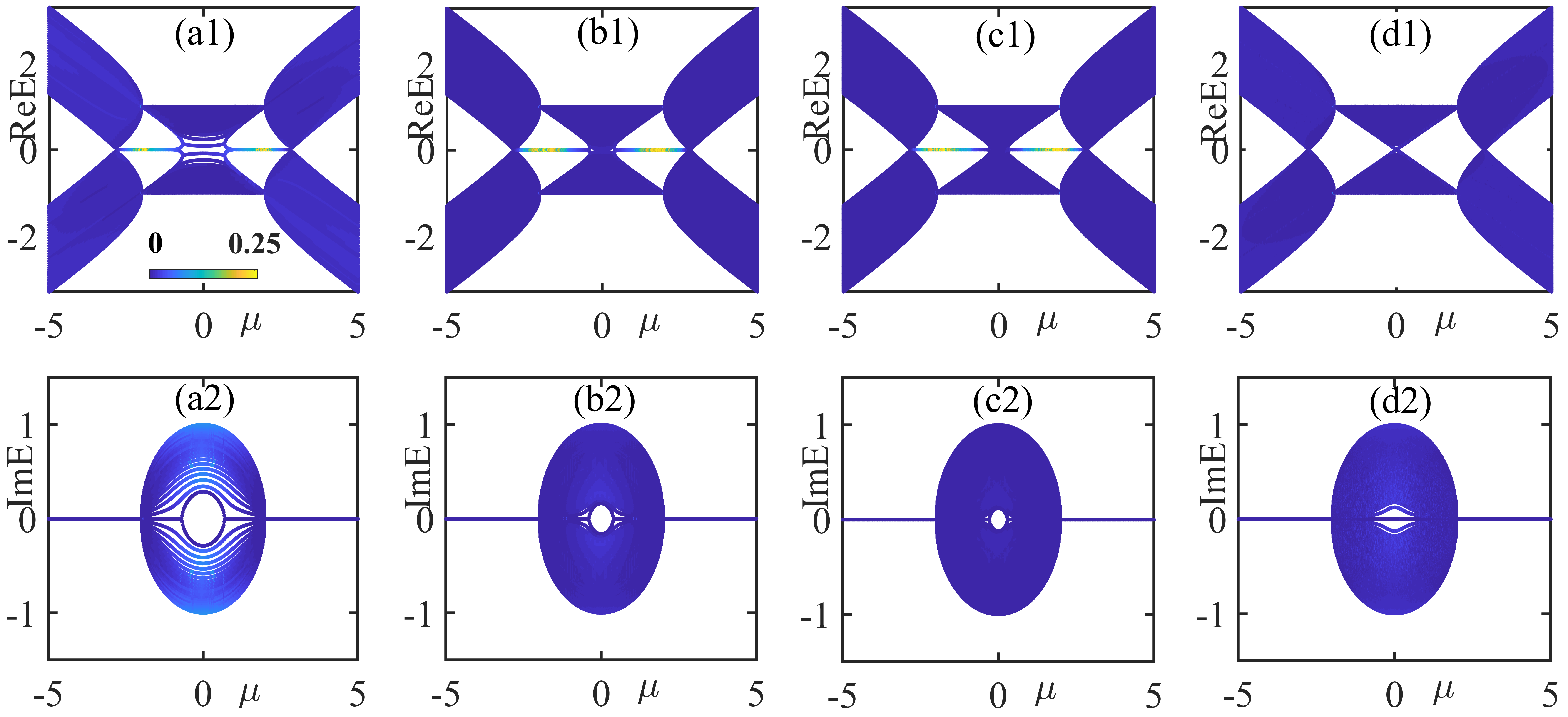}
	\caption{(Color online) Eigenenergy spectrum for the system with  $\gamma_1=2.0$ and $\gamma_2=0$ for different system size: (a) $L=100$, (b) $L=400$, and (c) $L=800$ under open boundary conditions. Panel (d) shows the corresponding spectrum under periodic boundary conditions. Other parameters are $t=1$, $\Delta=1$.}
	\label{fig3}
\end{figure}

At $\gamma_1=2t$, the real line gap further closes at $\mu=0$, and an imaginary line gap opens, as shown in Fig.~\ref{fig2}(b). In finite systems under open boundary conditions, the real part of the spectrum appears gapless in a small region around $\mu=0$ [Fig.~\ref{fig2}(b1)], which is a finite-size effect. In particular, near the gap-closing point, the OBC spectrum can deviate noticeably from the PBC spectrum. By increasing the system size, the OBC spectrum converges to the PBC result. In Fig.~\ref{fig3}, we present the eigenenergy spectra for different system sizes at $\gamma_1=2t$ and $\gamma_2=0$. As the system size increases from $L=100$ to $L=800$, the apparent gapless region in the real part of the spectrum around $\mu=0$ progressively shrinks, while the gap in the imaginary part near $\mu=0$ also decreases. This size dependence indicates that $\gamma_1=2t$ and $\gamma_2=0$ corresponds to a critical point at which the real line gap closes and an imaginary line gap begins to open at 
$\mu=0$. This behavior is consistent with the spectrum obtained under periodic boundary conditions, shown for comparison in Fig.~\ref{fig3}(d).

As the pairing imbalance $\gamma_1$ is further increased, the gapless region in the real  part of the spectrum expands, accompanied by a widening of the imaginary line gap, as shown in Fig.~\ref{fig2}(c). The imaginary line gap closes at
\begin{equation}
	\mu = \pm \sqrt{-4\Delta^2 + (\gamma_1 + \gamma_2)^2},
\end{equation}
which can be obtained from Eq.~(\ref{Ek}) by setting $k=\pi$ and requiring the imaginary part of the spectrum to vanish. For larger values of the chemical potential, $|\mu|>\sqrt{-4\Delta^2 + (\gamma_1 + \gamma_2)^2}$, the spectrum becomes gapped again in the real part. Similar spectral features are observed for nonzero $\gamma_2$, indicating that these real–imaginary gap transitions are generic in the presence of staggered pairing imbalance.

In addition, the inspection of the imaginary parts of the spectrum in Fig.~\ref{fig2} shows that, for sufficiently large $|\mu|$, the eigenenergies become purely real, signaling a complex-to-real transition in the spectrum. For the representative case $t=\Delta=1$, the transition occurs at $\mu=\pm |\gamma_1 -\gamma_2|$. Overall, the introduction of pairing imbalance in the non-Hermitian Kitaev chain leads to rich spectral behavior, including real–complex transitions and the conversion between real and imaginary line gaps. These features play a crucial role in determining the topological phases of the system, as discussed in the following section.

\section{Topological superconducting phase}\label{Sec4}
We now investigate the topological properties of the non-Hermitian Kitaev chain with staggered pairing imbalance. For this model, the non-Hermitian skin effect (NHSE) is not present, as can be seen from the IPR values of bulk states shown in Figs.~\ref{fig2} and \ref{fig3}, which are close to zero, indicating that the bulk states are extended. Different from the systems with nonreciprocal hopping, the pairing imbalances includes terms such as $\gamma c_j^\dagger c_{j+1}^\dagger$ or $-\gamma c_{j+1} c_j$, while the hopping amplitudes will always be reciprocal along the two directions in the lattice. This can further be illustrated by writing the model Hamiltonian in the Majorana fermion basis, where the 1D lattice is transformed into a ladder composed of Majorana fermions. The hopping between these fermions are reciprocal in the system, as will be shown in Fig.~\ref{fig6}. Thus, no NHSE is observed in our system. The OBC and PBC spectrum are thus consistent with each other. As a result, the phase boundaries of the topological superconducting phases can be determined from the gap-closing conditions of the bulk spectrum under periodic boundary conditions. 

As shown in Fig.~\ref{fig2}, Majorana zero modes appear inside the spectral gap in the topologically nontrivial regime. For $|\gamma_1|<2t$, the nontrivial phase exists when
\begin{equation}
	|\mu|<\sqrt{4t^2 + (\gamma_1-\gamma_2)^2}.
\end{equation}
Compared with the Hermitian Kitaev chain, where MZMs exist only for $|\mu|<2t$, the presence of pairing imbalance substantially enlarges the parameter region supporting a topological superconducting phase.

When $|\gamma_1|>2t$ and $|\Delta|<2t$, the topologically nontrivial phase is realized within the window
\begin{equation}\label{TopoPhase}
	\sqrt{-4\Delta^2 + (\gamma_1+\gamma_2)^2}<|\mu|<\sqrt{4t^2 + (\gamma_1-\gamma_2)^2},
\end{equation}
which follows from the gap-closing conditions at $k=0$ and $k=\pi$ in Eq.~(\ref{Ek}). Notably, if either $\gamma_1$ or $\gamma_2$ vanishes, or $\gamma_1$ and $\gamma_2$ are of opposite signs, a finite interval of $\mu$ satisfying Eq.~(\ref{TopoPhase}) always exists. This implies that a topologically nontrivial phase with MZMs can persist even for arbitrarily large chemical potential, in sharp contrast to the Hermitian Kitaev chain. In the following, we will investigate the properties of the edge modes and calculate the topological invariant characterizing the nontrivial phase.

\begin{figure}[t]
	\includegraphics[width=3.4in]{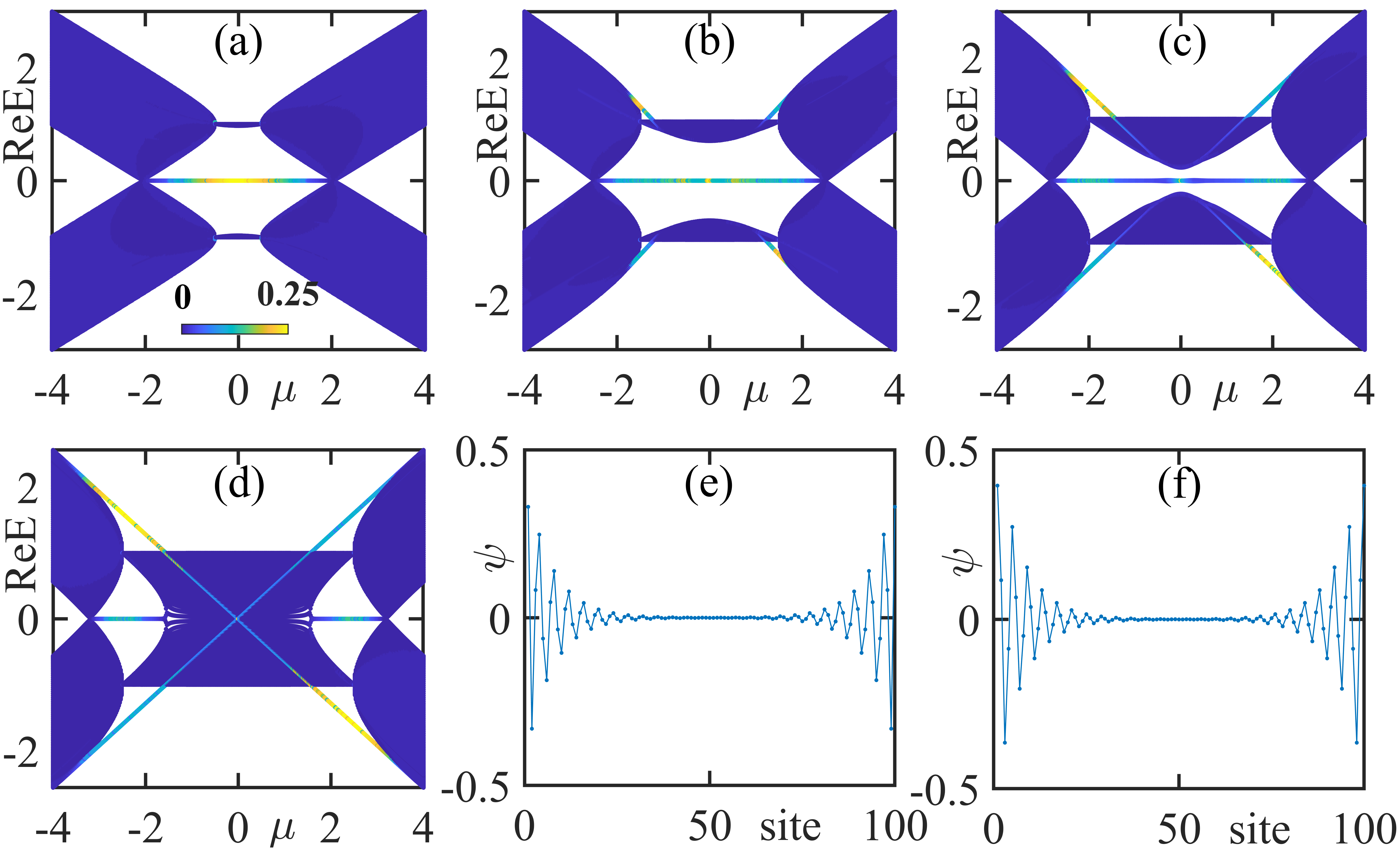}
	\caption{(Color online) Real and imaginary parts of the eigenenergy spectrum as a function of chemical potential $\mu$ for different values of $\gamma_2$ with $\gamma_1=0$. Panels (a)–(d) correspond to (a) $\gamma_2=0.5$, (b) $\gamma_2=1.5$, (c) $\gamma_2=2.0$, and (d) $\gamma_2=2.5$. Panels (e) and (f) show the spatial probability distributions of the Majorana zero modes and finite-energy Majorana edge modes, respectively. Other parameters are $t=1$, $\Delta=1$, and system size $L=2N=200$.}
	\label{fig4}
\end{figure}

\subsection{Coexistence of zero- and finite-energy Majorana edge modes}
For the case with $\gamma_1=0$ and $\gamma_2 \neq 0$, the parameter regions supporting MZMs are qualitatively similar, as shown in Fig.~\ref{fig4}. The spatial profiles of the MZMs are presented in Fig.~\ref{fig4}(e). In addition to the zero-energy modes, finite-energy Majorana edge modes emerge when $\gamma_2$ becomes sufficiently large, as illustrated in Fig.~\ref{fig4}(b). With increasing $\gamma_2$, these finite-energy edge modes extend over a broader parameter region and may even enter the bulk spectrum, as shown in Figs.~\ref{fig4}(c) and \ref{fig4}(d). In total, there are four finite-energy edge modes, with their spatial distributions shown in Fig.~\ref{fig4}(f). The corresponding eigenenergies depend approximately linearly on the chemical potential $\mu$. Again, note that in Fig.~\ref{fig4}(c), the gap should be closed at $\mu=0$, but for finite size system, the gap seems still open in this case. By increasing  the system size, the gap will become closed.

\begin{figure}[t]
	\includegraphics[width=3.2in]{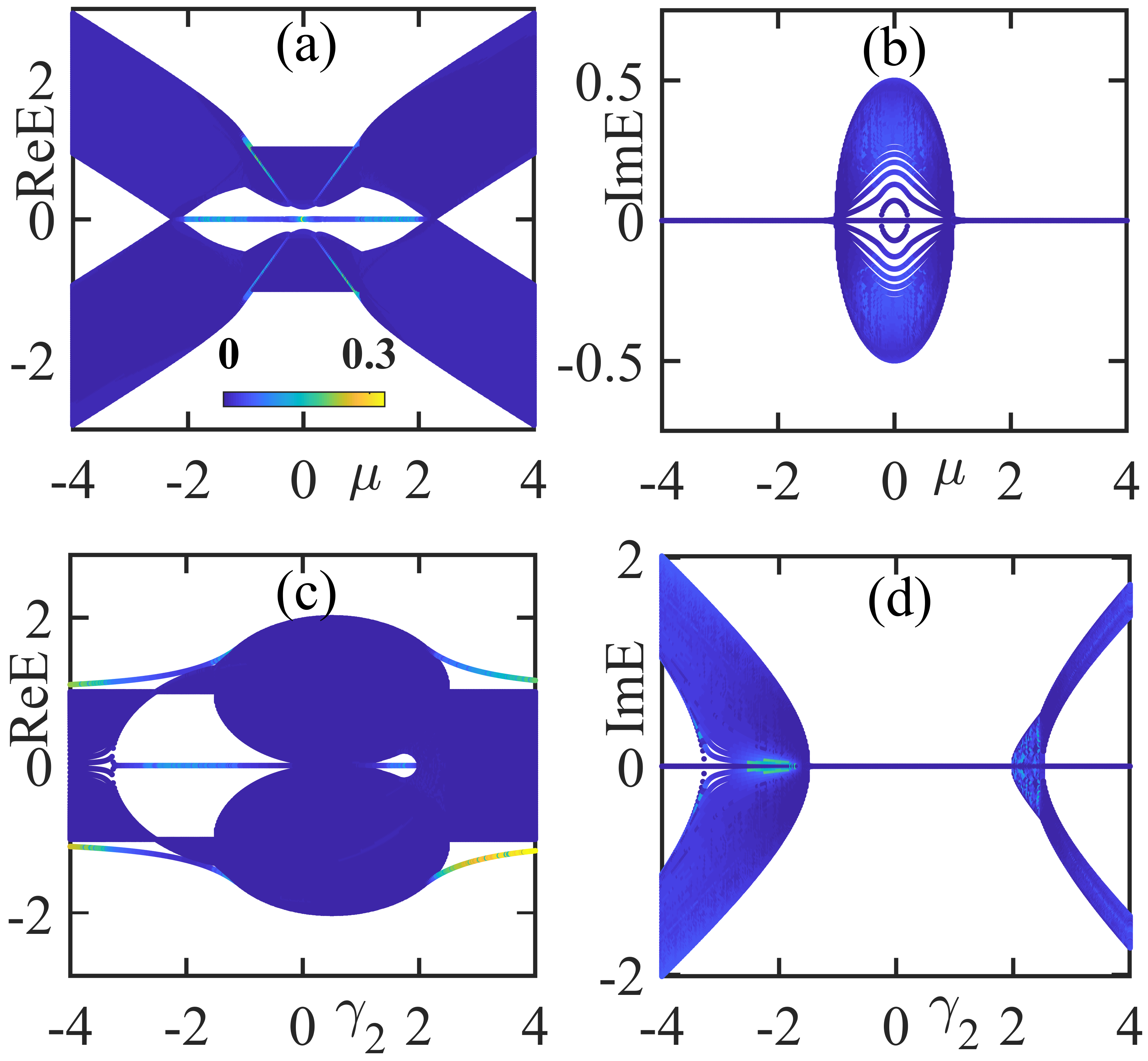}
	\caption{(Color online) Eigenenergy spectrum of the non-Hermitian Kitaev chain for finite pairing imbalance in both intra- and intercell pairing terms. Panels (a) and (b) show the real and imaginary parts of the spectrum, respectively, as functions of the chemical potential $\mu$ for $\gamma_1=0.5$ and $\gamma_2=1.5$. Panels (c) and (d) show the real and imaginary parts of the spectrum, respectively, as functions of the intercell pairing imbalance $\gamma_2$ for $\gamma_1=0.5$ and $\mu=2$. Other parameters: $t=\Delta=1$, $L=200$.}
	\label{fig5}
\end{figure}

When both $\gamma_1$ and $\gamma_2$ are nonzero, similar spectral and topological features are observed, as shown in Figs.~\ref{fig5}(a) and \ref{fig5}(b). The real part of the spectrum becomes gapless at $\mu=0$ when $|\gamma_1 + \gamma_2|=2t$, and MZMs appear for $|\mu|<\sqrt{4t^2 + (\gamma_1-\gamma_2)^2}$, consistent with Eq.~(\ref{TopoPhase}). For $|\mu|>|\gamma_1-\gamma_2|$, the spectrum becomes purely real, signaling a complex-to-real transition. Finite-energy edge modes can again be observed inside the bulk spectrum [Fig.~\ref{fig5}(a)].

Figures~\ref{fig5}(c) and \ref{fig5}(d) show the spectrum as a function of $\gamma_2$ for $\gamma_1=0.5$t and $\mu=2t$. In this case, the two topologically nontrivial regions are asymmetric with respect to $\gamma_2$, owing to the simultaneous presence of intra- and intercell pairing imbalance. As $\gamma_1$ increases, the nontrivial region at positive $\gamma_2$ gradually shrinks and eventually disappears, whereas the region at negative $\gamma_2$ remains robust. Finite-energy edge modes emerge when $|\gamma_2|$ becomes sufficiently large, and the spectrum undergoes a real–complex transition, as shown in Fig.~\ref{fig5}(d).

\begin{figure}[t]
	\includegraphics[width=3.4in]{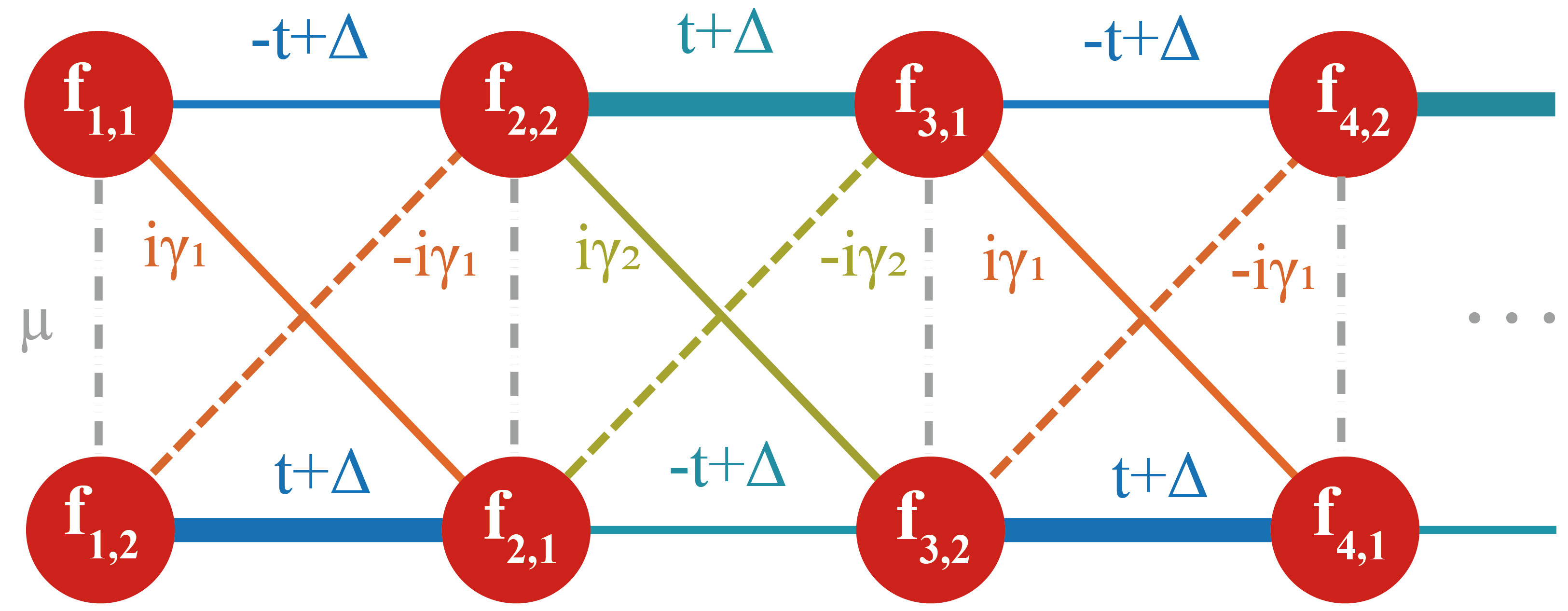}
	\caption{(Color online) Schematic illustration of the effective two-leg ladder representation of the one-dimensional Kitaev chain in the Majorana fermion basis. Red solid circles denote Majorana fermion operators, and the connecting lines indicate the couplings induced by chemical potential, hopping and imbalanced superconducting pairing terms.}
	\label{fig6}
\end{figure}

To gain further insight into the emergence of finite-energy Majorana edge modes, we rewrite the Hamiltonian in terms of Majorana operators by introducing $c_{j,A}=\frac{1}{2}(f_{2j-1,1}+i f_{2j-1,2})$ and $c_{j,B}=\frac{1}{2}(f_{2j,1}+i f_{2j,2})$, where the Majorana operators satisfy $\left\{f_{m,a}, f_{n,b} \right\} = 2 \delta_{m,n} \delta_{a,b}$ with $a,b \in \left\{1, 2 \right\}$. The Dirac fermion is now expressed as the combination of two Majorana fermions, denoted by $f_{2j-1,1}$ and $f_{2j-1,2}$ for the $A$ site, and $f_{2j,1}$ and $f_{2j,2}$ for the $B$ site in the $j$th unit cell, respectively. In this representation, the Kitaev chain maps onto a ladder model composed of Majorana fermions, as schematically illustrated in Fig.~\ref{fig6}. The chemical potential $\mu$ induces inter-leg couplings, while the hopping $t$ and SC pairing term $\Delta$ generate nearest-neighbor couplings along each leg. The staggered pairing imbalance further introduces diagonal couplings between Majorana fermions on opposite legs.

The interplay among these couplings leads to distinct patterns of Majorana dimerization, which in turn give rise to both zero- and finite-energy edge states. The ladder structure can be viewed as comprising four coupled Su–Schrieffer–Heeger (SSH)–type chains: the upper leg, the lower leg, and two zigzag chains formed by the $f_{j,1}$ and $f_{j,2}$ Majorana operators, respectively. In the absence of diagonal couplings, corresponding to vanishing or small pairing imbalance, one of the two legs can realize a nontrivial SSH phase for small $\mu$, depending on the sign of $\Delta$. For example, when $\Delta>0$, the upper leg is topologically nontrivial while the lower leg remains trivial, leaving the Majorana fermions $f_{1,1}$ and $f_{N,2}$ unpaired at the two ends of the system and giving rise to Majorana zero modes. As the pairing imbalance increases, particularly when $|\gamma_2|>|\gamma_1|$, both zigzag SSH chains become topologically nontrivial. In this regime, the Majorana modes $f_{1,1}$ and $f_{1,2}$ remain unpaired at the left end, while $f_{N,1}$ and $f_{N,2}$ remain unpaired at the right end. The finite inter-leg coupling $\mu$ hybridizes these edge Majorana modes locally, lifting their energies away from zero and generating finite-energy Majorana edge states. Therefore, this Majorana ladder picture provides an intuitive and unified explanation for the coexistence of Majorana zero modes and finite-energy edge modes observed in our system.

The finite-energy edge modes are not topologically protected in the same sense as the Majorana zero modes, and their eigenenergies vary continuously with the chemical potential and the pairing imbalance, as shown in Figs.~\ref{fig5}(a) and \ref{fig5}(c). In particular, the approximately linear dependence of these edge-mode energies on $\mu$ can be understood using the approach developed in Ref.~\cite{Zhang2025JPCM}. From the Majorana fermion ladder model shown in Fig.~\ref{fig6}, we can see that the finite-energy edge modes originate from the staggered diagonal pairing between the Majorana fermions and the coupling between the two legs, which are determined by the chemical potential and the staggered imbalance in the pairing terms. So, if the staggered imbalances are removed, there will be no finite-energy edge modes in the system. However, the Majorana zero modes reside in the energy gap and exist in a wide range of system parameters, which are robust to perturbations. Notably, in the present non-Hermitian system, Majorana zero modes and finite-energy edge modes can coexist within the same parameter regime. This behavior contrasts with the Hermitian Kitaev chain with staggered superconducting pairing studied in Ref.~\cite{Zhang2025JPCM}, where zero-energy and finite-energy edge modes appear in distinct regions of parameter space.

\subsection{Topological invariant and phase diagram}
As we have analyzed in Sec.~\ref{Sec2}, for the general cases with $\gamma_1 \neq \gamma_2$, the PHS in the model Hamiltonian is preserved and the model belongs to class $D^\dagger$ according to the classification introduced in Ref.~\cite{Kawabata2019PRX}. Since we have no point gap and no NHSE, the topological invariant will depend on whether the line gap in the spectrum is real or imaginary. As a result, the topology can be characterized by using a line-gap classification. When a real line gap is present, the system admits a well-defined $\mathbb{Z}_2$ topological invariant, which can be formulated in terms of the Pfaffian of the antisymmetric matrix at the high-symmetric momenta $k=0$ and $\pi$~\cite{Kitaev2001}. The resulting invariant correctly captures the emergence of Majorana zero modes and reproduces the analytically determined phase boundaries. However, if the energy gap is an imaginary gap, then the system is trivial and there will be no topological edge modes. Therefore, due to its non-Hermitian nature, the model studied here exhibits enriched spectral and boundary phenomena due to pairing imbalance.

In the present model, the BdG Hamiltonian can be cast into a skew-symmetric form, allowing the Pfaffian to be evaluated at high-symmetry momenta. Specifically, at the $k=0,\pi$, the Hamiltonian obeys $h(k)=-\tau_x h^{T}(k) \tau_x$, which allows it to be brought into an antisymmetric form. Here $\tau_x$ is the Pauli matrix. Defining $A(k)= ih(k) \tau_x$, one finds that $A(k)$ satisfies $A(k)=-A^T(k)$ for $k=0, \pi$, independent of whether $h(k)$ is Hermitian. Substituting $h(k)$ in Eq.~\ref{hk} into $A(k)$, we have
\begin{widetext}
\begin{equation}
		A(k)=i \begin{bmatrix} 
			0 & - (\Delta +\gamma_1)+(\Delta +\gamma_2) e^{-ik} & -\mu & -t(1+e^{-ik})   \\
			(\Delta +\gamma_1) - (\Delta +\gamma_2) e^{ik} & 0 & -t(1+e^{ik}) & -\mu \\
			\mu & t(1+e^{-ik}) & 0 & (\Delta -\gamma_1) - (\Delta -\gamma_2) e^{-ik}  \\
		    t(1+e^{ik}) & \mu & - (\Delta -\gamma_1)+(\Delta -\gamma_2) e^{ik} & 0 
		\end{bmatrix}.
\end{equation} 
\end{widetext}
which is manifestly skew-symmetric at $k=0,\pi$. The Pfaffian $Pf[A(k)]$ is in general complex due to non-Hermiticity. However, as long as the system possesses a real line gap, the Pfaffian phase can be chosen continuously across the Brillouin zone, allowing the definition of a $\mathbb{Z}_2$ topological invariant
\begin{equation}
	\nu = \text{sgn} \left\{ Pf[A(0)]Pf[A(\pi)] \right\}.
\end{equation}
A change in $\nu$ coincides with a bulk gap closing at $k=0$ or $k=\pi$, consistent with the analytical phase boundaries obtained from the bulk dispersion. From the explicit expressions for $Pf[A(0)]$ and $Pf[A(\pi)]$, we can find that the nontrivial phase $(\nu=-1)$ and trivial phase $(\nu=1)$ are separated by the critical lines
\begin{equation}
	\left\{
	\begin{array}{ll}
		\mu^2 &=  4t^2 + (\gamma_1-\gamma_2)^2, \\
		\mu^2 &=  -4\Delta^2 + (\gamma_1+\gamma_2)^2,
	\end{array}
	\right.
\end{equation} 
which are consistent with the gap closing points obtained from the energy spectrum in momentum space. The two critical lines intersect only when the condition $t^2+\Delta^2=\gamma_1 \gamma_2$ is satisfied. Consequently, for $\gamma_1 \gamma_2 \leq 0$, the two phase boundaries never cross, ensuring the persistence of a finite topologically nontrivial region in parameter space.

\begin{figure}[t]
	\includegraphics[width=3.4in]{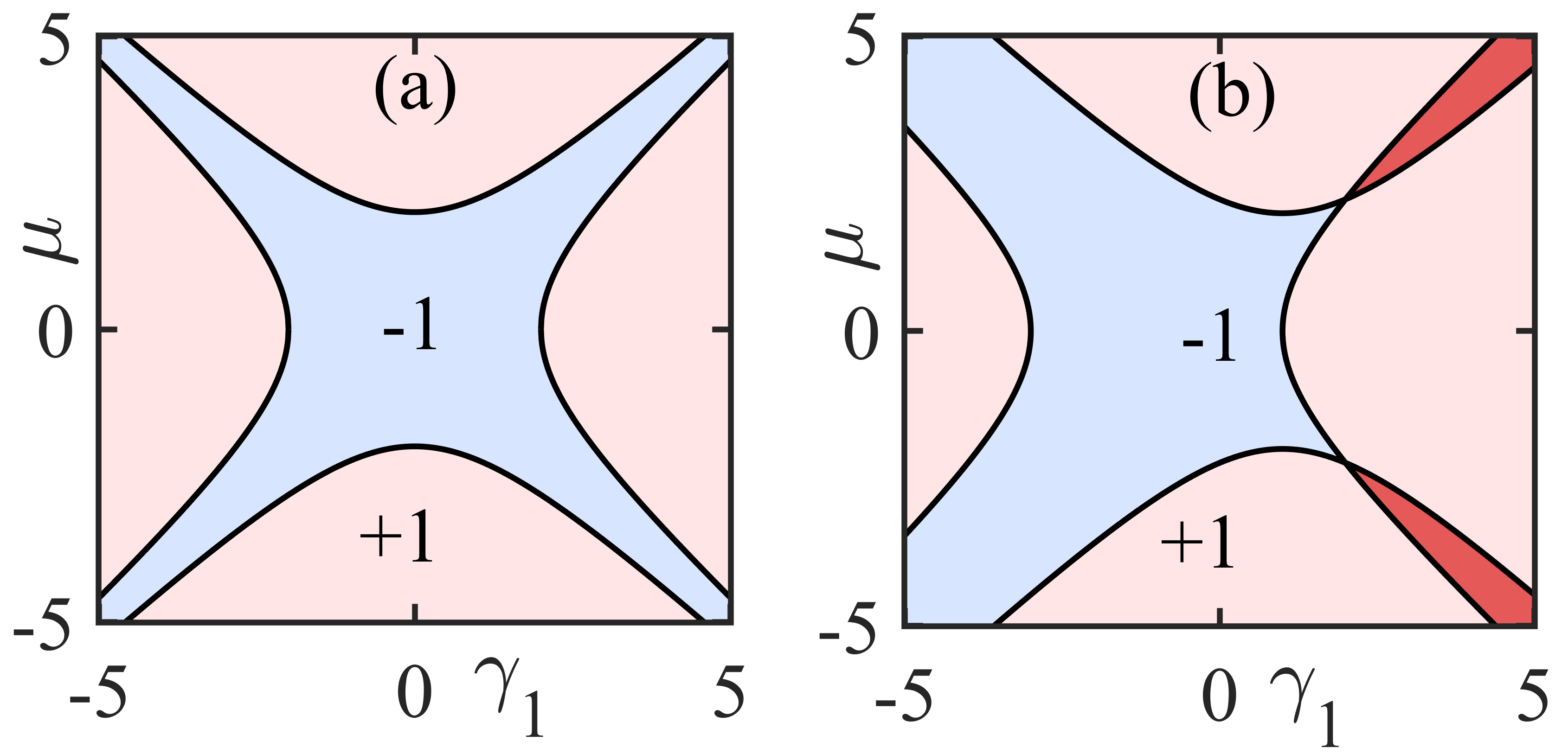}
	\caption{(Color online) Topological phase diagram of the non-Hermitian Kitaev chain in the $\mu-\gamma_1$ plane for (a) $\gamma_2=0$ and (b) $\gamma_2=1$. The blue (pink) regions denote the topologically nontrivial (trivial) phase, with $-1$ ($+1$) being the corresponding  $Z_2$ topological invariant. The red region in (b) is also trivial as the band is separated by imaginary line gap and there are no MZMs in the system. Other parameters are $t=1$ and $\Delta=1$.}
	\label{fig7}
\end{figure}

Figure~\ref{fig7} presents the phase diagrams in the $\mu-\gamma_1$ plane for $\gamma_2=0$ and $\gamma_2=1$. From the phase diagram, we can see that as long as $\gamma_1 \gamma_2 \leq 0$—namely, when one pairing imbalance vanishes or the two imbalances have opposite signs—the topologically nontrivial phase always survives, regardless of the strength of the chemical potential or pairing imbalance. The interplay between chemical potential and staggered pairing imbalance therefore stabilizes robust Majorana zero modes in a broad parameter regime. We note that in the red region of Fig.~\ref{fig7}(b), the system is topologically trivial despite yielding $\nu=-1$ from the Pfaffian evaluation. In this parameter regime, $|\mu|< \sqrt{-4\Delta^2 + (\gamma_1+\gamma_2)^2}$, the real part of the spectrum is gapless while the imaginary part remains gapped, which means that the energy gap is an imaginary line gap. Thus, the system is not characterized by the $\mathbb{Z}_2$  topological invariant. Instead, it is trivial according to the topological classification in Ref.~\cite{Kawabata2019PRX}, and no MZMs present in this situation under open boundary conditions.

It will be interesting to investigate the influences of disorders on this non-Hermitian Kitaev chain with staggered pairing imbalances. From the phase diagrams shown in Fig.~\ref{fig7}, we can find that the topologically nontrivial region in the $\mu-\gamma_1$ plane shrinks gradually when $\mu$ and $\gamma_1$ increases. So, for moderate disorders in the system, we can believe that the nontrivial phase can persist if $\mu$ and $\gamma_1$ are not too large. However, whether the nontrivial region still appears in the disordered system with large $\mu$ need further investigations, which will be explored in future studies.

\section{Summary}\label{Sec5}
In summary, we have studied a one-dimensional non-Hermitian Kitaev chain with staggered imbalances in the $p$-wave superconducting pairing, which effectively introduce gain and loss into the pairing processes. By tuning the chemical potential and the pairing imbalance, we demonstrated that the system exhibits rich non-Hermitian spectral features, including real–complex transitions and the conversion between real and imaginary line gaps. The gap-closing points and transition boundaries were analytically determined from the momentum-space spectrum, and the numerical results under open boundary conditions were shown to be consistent with the bulk analysis. We further revealed that the introduction of pairing imbalance can significantly enhance the topological superconducting phase compared with the Hermitian Kitaev chain. In particular, the parameter region supporting Majorana zero modes is enlarged, and, remarkably, topologically nontrivial phases can persist even for arbitrarily large chemical potential when the pairing imbalance is staggered appropriately. The topological phase boundaries were characterized by a $\mathbb{Z}_2$ invariant defined via the Pfaffian of the BdG Hamiltonian. In addition to Majorana zero modes, we also identified the coexistence of zero-energy and nonzero-energy Majorana edge modes in certain parameter regimes. These finite-energy edge states emerge as a direct consequence of the staggered non-Hermitian pairing and can persist even inside the bulk spectrum. By expressing the model in terms of Majorana fermion operators, we provided an intuitive picture for the origin of these edge modes in terms of effective couplings in a Majorana ladder representation.

Compared with the Hermitian Kitaev chain with staggered superconducting pairing in Ref.~\cite{Zhang2025JPCM}, the results obtained in this work show that instead of destroying the topological phase, the non-Hermitian pairing imbalance can enlarge the nontrivial regions. The interplay between the chemical potential and staggered pairing imbalance leads to a region host nontrivial phase even when the chemical potential becomes extremely large, which is absent in the Hermitian case. Moreover, the influences of the real/imaginary line gaps on the topological phase further exhibits the exotic features of non-Hermitian systems.

Experimentally, the staggered superconducting pairing in the Hermitian case has already been proposed by proximitizing a one-dimensional semiconductor with a periodic array of superconductors, as proposed in Refs.~\cite{Levine2017PRB,Escribano2019PRB}. The spatially periodic deposition of superconducting islands naturally induces a staggered modulation of the pairing amplitude along the wire. Alternatively, similar model may also be implemented in quantum-dot-based platforms, where Majorana-like states have been observed in few-dot systems~\cite{Dvir2023Nature,Haaf2024Nature,Zatelli2024NatCom,Bordin2025NatNano}. By further introducing imbalances in the pairing terms, our model might also been realized in these setups, allowing access to both zero-energy and finite-energy edge modes predicted here. The coexistence of these edge modes provides a setting to study non-Hermitian effects in topological superconductors and their influence on spectral and transport properties.

Our results demonstrate that non-Hermitian pairing imbalance can play a constructive role in stabilizing and enhancing topological superconducting phases, rather than merely destroying them. This work uncovers non-Hermitian topological phenomena in superconducting systems and provides a platform for exploring Majorana physics and topological superconductivity beyond the Hermitian paradigm.


\begin{acknowledgments}
This work is supported by the National Natural Science Foundation of China under Grant No. 12204326. R.L. is supported by the Quantum Science and Technology-National Science and Technology Major Project (Grant No. 2021ZD0302100) and the 'Gravitational Wave Detection' program (2023YFC2205800) funded by the Ministry of Science and Technology of the People's Republic of China.
\end{acknowledgments}

\end{document}